\documentclass{ws-p8-50x6-00}
\usepackage{graphicx}

\newcommand\si{S_{\rm I}}
\newcommand\ti{T_{\rm I}}
\newcommand\sr{S_{\rm R}}
\newcommand\tr{T_{\rm R}}
\begin{document}


\begin{center}
{\bf \LARGE Cosmological Aspects of Heterotic M-theory} \\
\vspace{1.5cm}
T. Barreiro\footnote{e-mail address: mppg6@pcss.maps.susx.ac.uk} \\
{\em Centre for Theoretical Physics, University of Sussex \\
Falmer, Brighton BN1 9QJ, UK} \\
\vspace{0.5cm}
B. de Carlos\footnote{e-mail address: Beatriz.de.Carlos@cern.ch} \\
{\em Theory Division, CERN, CH-1211 Geneva 23, Switzerland} \\
\end{center}
\vspace{1.7cm}
\noindent
In this talk we discuss a few relevant aspects of heterotic M-theory. 
These are the stabilisation of the two relevant moduli (the length of the 
eleventh segment ($\pi \rho$) and the volume of the internal six manifold 
($V$)) in models where supersymmetry is broken by multiple gaugino 
condensation and non-perturbative corrections to the K\"ahler potential; 
the existence of almost flat directions in the scalar potential; the 
possibility of lifting them, and their role in constructing a viable model 
of inflation. Finally, we review the status of the moduli problem within 
these models.

\vspace{1.2cm}

\begin{center}
{Talk given at the \\ {\bf  International Workshop on Particle Physics and 
the Early Universe (COSMO-99)} \\
Trieste, Italy, September 27$^{th}$ - October 2$^{nd}$, 1999} \\
{(\em to be published in the proceedings)}\\
\end{center}

\thispagestyle{empty}

\vspace{1cm} 
\noindent
CERN-TH/2000-023 \\
SUSX-TH/00-001 \\
January 2000

\vskip3in

\newpage

\title{Cosmological Aspects of Heterotic M-theory}

\author{T. Barreiro}

\address{Centre for Theoretical Physics, University of Sussex \\
Falmer, Brighton BN1 9QJ, UK\\
E-mail: mppg6@pcss.maps.susx.ac.uk}  

\author{B. de Carlos}

\address{Theory Division, CERN, CH-1211 Geneva 23, Switzerland \\
E-mail: Beatriz.de.Carlos@cern.ch}


\maketitle

\abstracts{In this talk we discuss a few relevant aspects of 
heterotic M-theory. These are the stabilization of the two relevant
moduli (the length of the eleventh segment ($\pi \rho$) and the volume 
of the internal six manifold ($V$)) in models where supersymmetry is 
broken by multiple gaugino condensation and non-perturbative corrections
to the K\"ahler potential; the existence of almost flat directions in the
scalar potential; the possibility of lifting them, and their role in 
constructing a viable model of inflation. Finally, we review the status 
of the moduli problem within these models.}

\section{Introduction}

As it was well established by Ho\v{r}ava and Witten~\cite{Horav96}, the
strong coupling limit of the $E_8 \times E_8$ heterotic string theory 
can be described by $d=11$ supergravity (SUGRA). This theory can be 
compactified on a manifold with boundaries, usually expressed as
$X \times S^1/Z_2$, where $X$ is the $d=6$ Calabi--Yau manifold and 
$S^1/Z_2$ is the so-called eleventh segment. Altogether the picture we 
get is that of two walls (so-called hidden and observable) that interact
through gravity.

The relevant parameters in this theory are $V$, the volume of the $d=6$ 
manifold, and $\pi \rho$, the length of the eleventh segment. Among the 
attractive features of this theory, it is worth mentioning the fact that 
gauge coupling unification is now entirely natural~\cite{Witte96}, and the 
scale at which the gauge couplings meet, $10^{16}$ GeV, is reconciled with 
the $d=4$ Planck scale, $M_{\rm P} \sim 10^{18}$ GeV. For this to happen, 
the process of compactification must occur in the order $d=11 \rightarrow 
d=5 \rightarrow d=4$. In other words, in ordet to fit the 
phenomenologically preferred values for $\alpha_{\rm GUT}$, $M_{\rm GUT}$ 
and $M_{\rm P}$, given by
\begin{eqnarray}
\alpha_{\rm GUT} & = & (4 \pi)^{2/3} \kappa^{4/3} \langle V \rangle^{-1}
\; \; ,\nonumber \\ 
M_{\rm GUT} & = &  \langle V \rangle^{-1/6} \; \; ,\label{guts} \\
M_{\rm P} & = & \kappa^{-1} \sqrt{\pi \rho V} \;\;,\nonumber
\end{eqnarray}
we need $ \langle \pi \rho \rangle \sim  (4 \times 10^{15} \; 
{\rm GeV})^{-1}$ and $\langle V \rangle  \sim  (3 \times 10^{16} \; 
{\rm GeV})^{-6}$. $\kappa^2$ is the $d=11$ gravitational coupling and its 
corresponding Planck scale is given by $M_{11} = \kappa^{-2/9}$.

Within this framework we are interested in studying phenomena such as
supersymmetry (SUSY) breaking and the cosmological evolution of the
moduli fields, which occur at scales below the compactification one. 
Therefore we shall concentrate on the $d=4$ effective SUGRA theory
from now on.

\section{Moduli stabilization}

We shall discuss the stabilization of moduli as a previous step to
address the question of their cosmological evolution. In order to do 
that, let us notice that the values of $V$ and $\pi \rho$ are determined 
by the chiral superfields $S$ (the dilaton) and $T$ (the modulus).
More precisely the real parts of these chiral fields are given by
\begin{eqnarray}
\sr & \sim & \kappa^{4/3} V \sim O(\alpha_{\rm GUT}^{-1}) \;\;,
\label{mod} \\
\tr & \sim & \kappa^{-2/3} \pi \rho V^{1/3} \sim O(\alpha_{\rm GUT}^{-1}) \;\;.
\nonumber
\end{eqnarray}
Note that, in the weakly coupled heterotic case, $\sr \sim 
O(\alpha_{\rm GUT}^{-1}) \sim O(20)$, the same as here, whereas $\tr \sim 
O(1)$ (always in $M_{\rm P}$ units).

To study the dynamical behaviour of these fields, and whether they acquire 
the desired vacuum expectacion values (VEVs), we analyse the scalar potential
\begin{equation}
V = e^K \left\{ (W_i+ K_i W) (K_i^j)^{-1} (\bar{W}^j + K^j \bar{W} ) - 3 
|W|^2 \right\} \;\;, 
\label{pot}
\end{equation}
where $W(S,T)$ is the superpotential and $K(\sr,\tr)$ is the K\"ahler 
potential. The sub (super) indices indicate derivatives of the functions
with respect to the (conjugate) fields. Together with the gauge kinetic 
functions, $f_a$, $W$ and $K$ determine the SUGRA Lagrangian.

To be more precise, we shall assume a particular source for SUSY breaking, 
which is gaugino condensation in the hidden wall~\cite{Horav96p}. This is, 
so far, the most promising mechanism for breaking SUSY at the right scale 
in order to give rise to an acceptable phenomenology. It assumes the 
existence of a strong-type interaction in the hidden sector of the theory 
which, below a certain scale $\Lambda$, triggers the formation of gaugino 
condensates, $\lambda \lambda$, and the breakdown of SUSY. More precisely, 
\begin{equation}
W \sim \langle \lambda \lambda \rangle \sim \sum_{i=1}^N C_i 
e^{-\alpha_i f_i} \;\;,
\label{sup}
\end{equation}
where the sum runs over all the condensing groups in the hidden sector
($G_H = G_1 \times G_2 \times \ldots \times G_N$), $\alpha_i$ are proportional
to the 1-loop $\beta$-function coefficients associated to each condensing 
group, $C_i$ are also related to each group's characteristics and $f_i$ 
are the already-mentioned gauge kinetic functions. In particular, in
the hidden sector $f_i = (S - n_i T/2)/4 \pi$, with $n_i$ being model
dependent coefficients and, in the observable wall, 
$f_{\rm obs}=(S+n_{\rm obs} T/2)/4 \pi$.

The K\"ahler potential is given by the expression
\begin{equation}
K = K_0 + K_{\rm np} \;\;,
\label{kahler}
\end{equation}
where $K_0 = - \ln(2 \sr) - 3 \ln(2 \tr)$ is the tree level piece and 
$K_{\rm np}$ stands for M-theoretic non-perturbative effects. For the latter
we will use a specific ansatz~\cite{Barre98}.

Now that we have defined all the functions we need to proceed with the 
analysis of the scalar potential, Eq.~(\ref{pot}). We have first of all 
confirmed the results of Choi et al.~\cite{Choi99} (obtained with a 
slightly different ansatz for $K_{\rm np}$), namely that for one condensate 
only the weakly coupled minimum exists. With two condensates ($W= C_1 
e^{-\alpha_1 f_1} +  C_2 e^{-\alpha_2 f_2}$), however, things become more 
interesting. It is possible to find plenty of examples of condensing groups 
for which both moduli are fixed at the desired VEVs and SUSY is broken at 
the right scale (i.e. the gravitino mass, $m_{3/2}$, is of order 1 TeV). In 
fact, in order to understand the vacuum structure a bit better, it is 
convenient to redefine the fields $S$ and $T$ as follows:

$\bullet$ The combinations
\begin{eqnarray}
\Phi_{\rm r}^- & \equiv & \sr - \frac{n_1 \alpha_1 - n_2 \alpha_2}
{2(\alpha_1 - \alpha_2)} \tr  \; , \label{phi-} \\
\Phi_{\rm i}^- & \equiv & \si - \frac{n_1 \alpha_1 - n_2 \alpha_2}
{2(\alpha_1 - \alpha_2)} \ti = \frac{4 \pi^2 k}{\alpha_1-\alpha_2}
\nonumber
\end{eqnarray} 
are fixed by the interplay between condensates, very much in the same 
way as the racetrack mechanism worked in the weakly coupled heterotic case
(note, from the second equation, that the condensates at the minimum, i.
e. for $k$ odd, are in opposite phase).

$\bullet$ The orthogonal combinations are potentially {\em flat}
\begin{eqnarray}
\Phi_{\rm r}^+ & \equiv & \frac{n_1 \alpha_1 - n_2 \alpha_2}
{2(\alpha_1 - \alpha_2)} \sr + \tr \; , \label{phi+} \\
\Phi_{\rm i}^+ & \equiv & \frac{n_1 \alpha_1 - n_2 \alpha_2}
{2(\alpha_1 - \alpha_2)} \si + \ti \;\;. \nonumber
\end{eqnarray} 
In fact we can easily check that these combinations of the fields do
not appear in the superpotential. Therefore the potential flatness of
$\Phi_{\rm r}^+$ will be lifted by the presence of the K\"ahler potential,
which depends on $\sr$, $\tr$, whereas $\Phi_{\rm i}^+$ remains totally
flat. This is a completely new feature associated to M-theory models,
which depends entirely on the structure of the gauge kinetic functions 
$f_a$.

Let us proceed to analyse cases with more than two condensates. There the
definitions of $\Phi_{\rm r,i}^-$ and $\Phi_{\rm r,i}^+$ can be easily 
generalized to many condensates, and it is also possible to show that the 
flat direction will only remain so if and only if all the $n_i$ coefficients 
that enter the definition of the gauge kinetic functions, $f_i$, are 
the same. This is shown in Fig.~1 where we plot the scalar potential 
as a function of $\Phi_{\rm i}^{\pm}$, for $\Phi_{\rm r}^{\pm}$ fixed, in
the case of two (Fig.~1a) and three (Fig.~1b) condensates. 
\begin{figure}[t]
\parbox{28pc}{
\includegraphics[width=14pc]{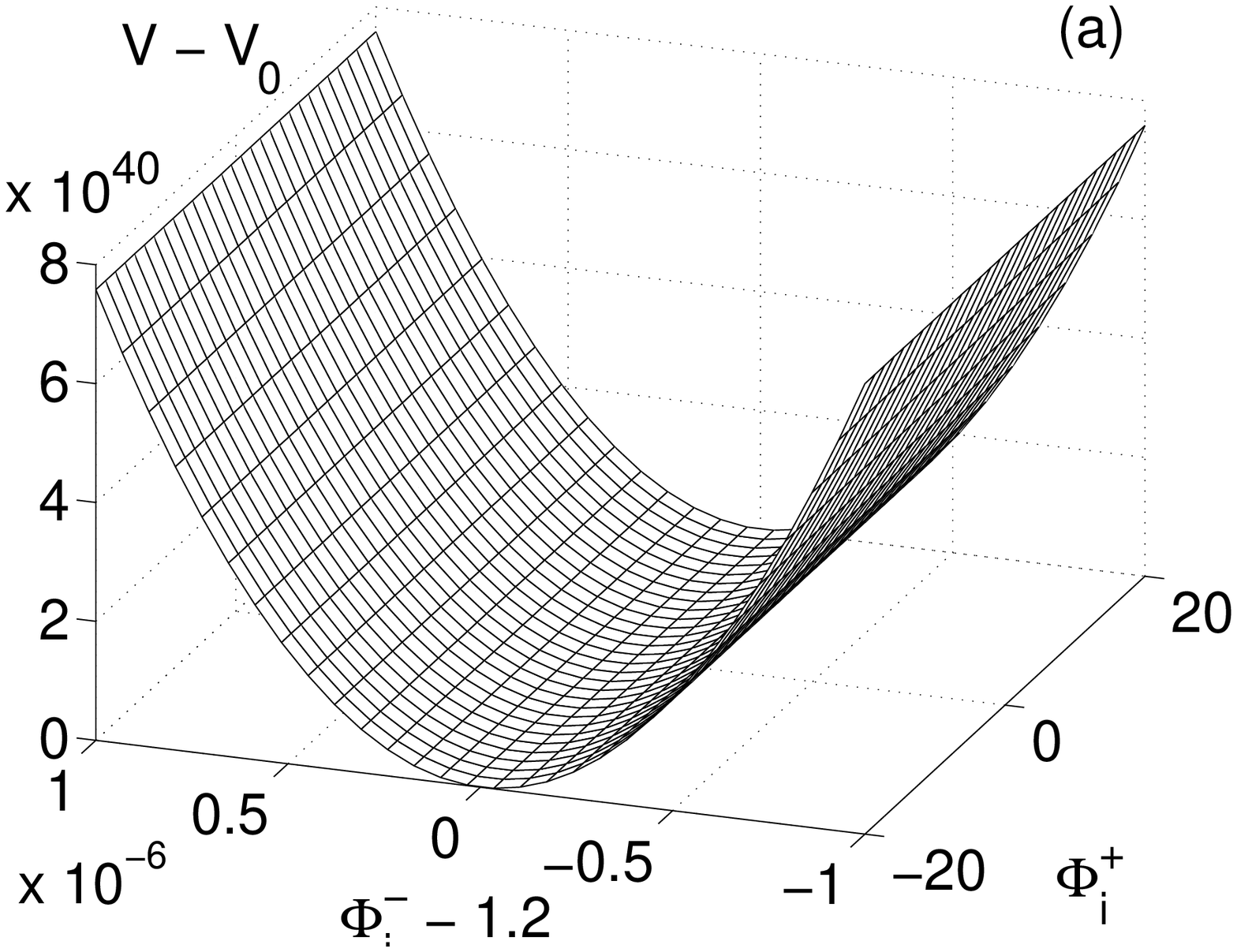}
\hfill
\includegraphics[width=14pc]{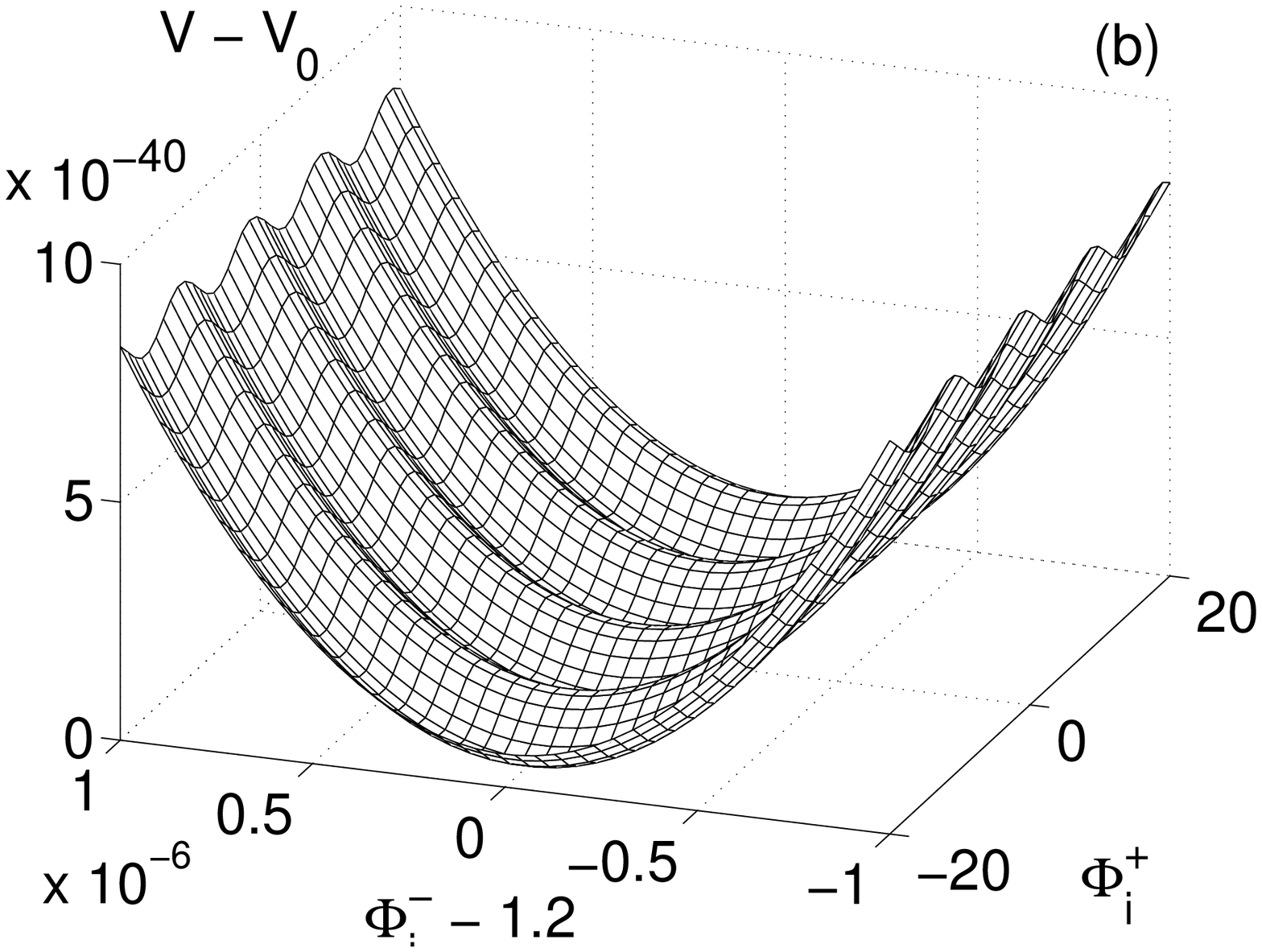}
}
\caption{Scalar potential for
(a) two condensates ${\rm SU}(3)_{M=0} \times {\rm SU}(4)_{M=8}$ with
$n_1 = n_2 = 1$, and
(b) an additional third condensate ${\rm SU}(2)_{M=0}$ with
$n_3 = 0.5$. $V_0$ is the value of the potential at the minimum.
}
\end{figure}
As we can see, the flat direction of Fig.~1a is lifted by the presence of 
a third condensate with $n_3 \neq n_1=n_2$. This opens up new scenarios 
from the cosmological point of view, as we are about to see.

\section{Cosmological Evolution and Moduli Problem}

Let us then study the possible cosmological role of
the $\Phi^{\pm}_{\rm r,i}$ fields.

$\bullet$ $\Phi^-_{\rm r}$ has an exponential-type potential. In fact it is
the analogous of the dilaton $S_{\rm R}$ in the weakly coupled heterotic 
string and will therefore suffer from the same problems~\cite{Brust93}.
The potential is too steep just before the minimum 
and the field tends to roll past it towards infinity. It can be, however,
stabilized in the presence of a dominating background~\cite{Barre98p}
but, in any case, it is totally unsuitable as an inflaton.

$\bullet$ $\Phi^-_{\rm i}$ has a steep sinusoidal potential, analogous to 
that of $S_{\rm I}$ in the weakly coupled heterotic case. Again, this field 
is not a good candidate for an inflaton.

On the other hand, the fields $\Phi_{\rm r,i}^+$ have more promising 
potentials, as it was pointed out above. It is then worth studying their
evolution equations in the presence of an expanding Universe
\begin{eqnarray}
K_{S \bar{S}} ( \ddot{S} + 3 H \dot{S} ) + K_{S S \bar{S}} \dot{S}^2 + 
\frac{\partial V}{\partial \bar{S}} = 0 \; \; , \label{evol} \\
K_{T \bar{T}} ( \ddot{T} + 3 H \dot{T} ) + K_{T T \bar{T}} \dot{T}^2 + 
\frac{\partial V}{\partial \bar{T}} = 0 \;\;,
\nonumber
\end{eqnarray}
where $H^2= \frac{1}{3} K_{S \bar{S}} \dot{S} \dot{\bar{S}} + \frac{1}{3} 
K_{T \bar{T}} \dot{T} \dot{\bar{T}} + \frac{V}{3}$ is the Hubble constant.
It is crucial to note here that both $S$ and $T$ have non-minimal kinetic 
terms, i.e. ${\cal L}_{kin} = K_{S \bar{S}} D_{\mu} S D^{\mu} \bar{S} + 
K_{T \bar{T}} D_{\mu} T D^{\mu} \bar{T}$. This is what introduces the new 
terms in the evolution equations (\ref{evol}).

Let us start with the $\Phi_{\rm r}^+$ direction. It is easy to show that 
the potential along this direction behaves as an inverse power-law, i.e.
$V(\Phi_{\rm r}^+) = A/(\Phi_{\rm r}^+)^n$, at least along the slope to the 
left of the minimum, which is the relevant one in terms of the evolution. By 
solving the evolution equations we have checked that the field inflates 
for a few e-folds before reaching its minimum. These kinds of inflationary 
scenarios are denoted as `intermediate' inflation in the 
literature~\cite{Musli90}. An important thing to notice is that, if the 
field $\Phi_{\rm r}^+$ were canonically normalized, then this inverse 
power-law potential would have given enough e-folds of inflation. 

\begin{figure}[t]
\parbox{28pc}{
\includegraphics[width=13pc]{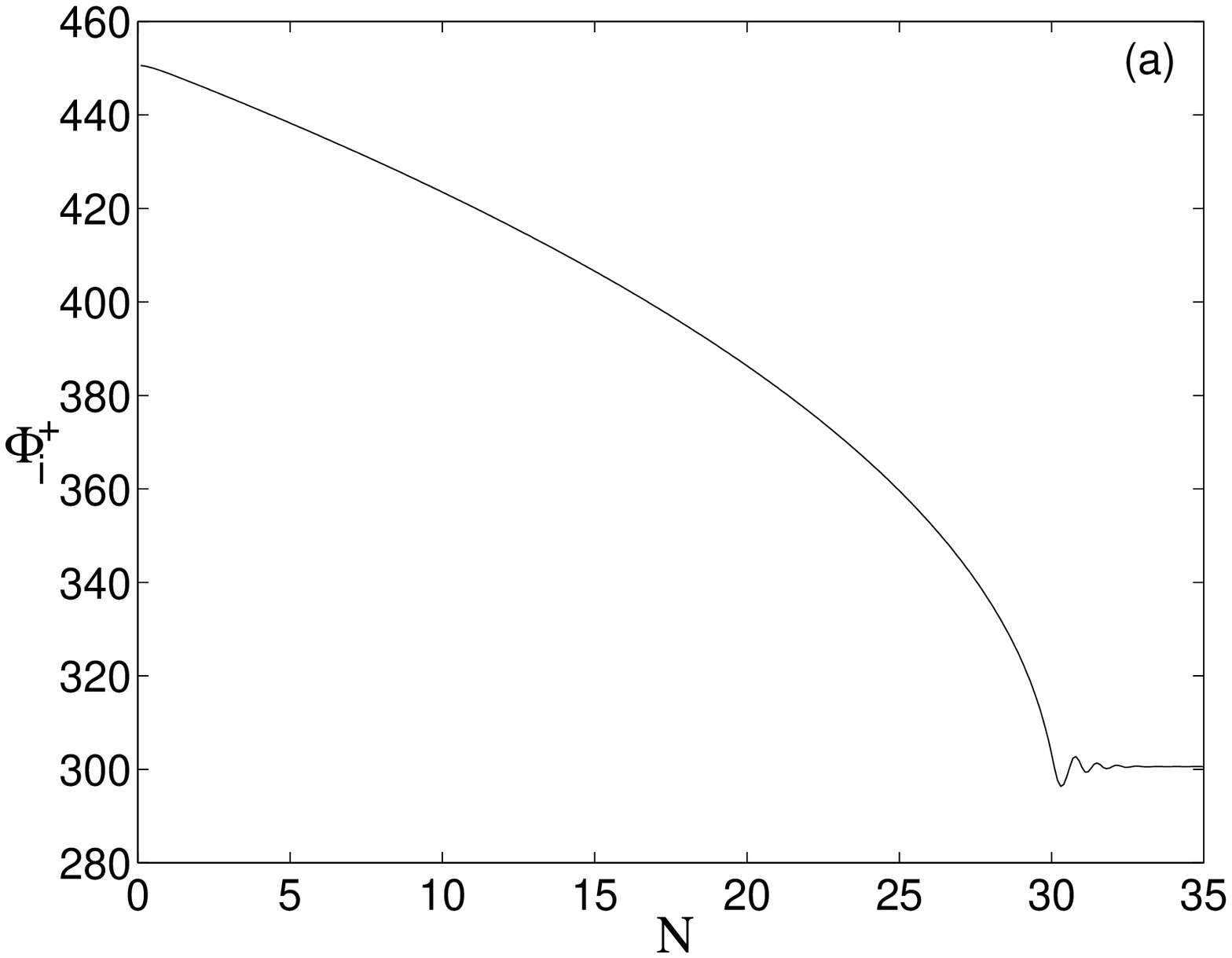}
\hfill
\includegraphics[width=13pc]{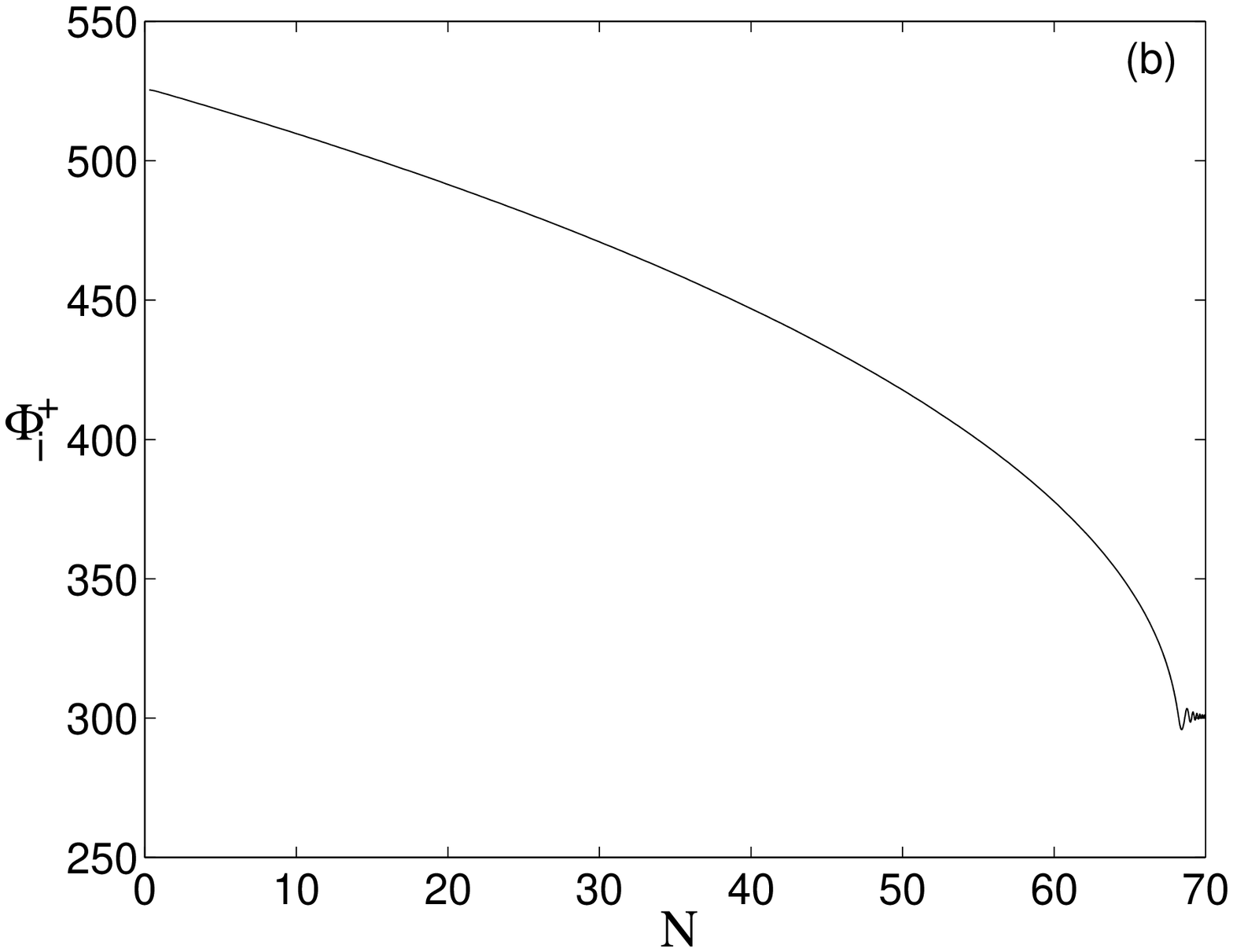}
}
\caption{Evolution of $\Phi^{+}_{\rm i}$ as a function of $N$ (number of 
e-folds) for the three condensates of Figure~1b, with $n3 = 0.995$, and two 
different initial conditions. The real fields are fixed to their minimum 
values.
}
\end{figure}

Finally we turn to the remaining direction to be analysed, namely 
$\Phi_{\rm i}^+$: as we had said before, in the two-condensate case this
is a totally flat direction, however when we introduce a third condensate
with $n_3 \neq n_1=n_2$, things change. It can be shown that the potential 
in this latter case is a sinusoidal one, 
\begin{equation}
V = V_0 [1+ \cos(a \Phi_{\rm i}^++b)] \;\;,
\label{Vim}
\end{equation}
where $a$ and $b$ are functions of $n_3$. It is easy to see that the closer
$n_3$ is to $n_1$ and $n_2$ the flatter the potential is. Therefore we have 
a way of controlling the flatness of the potential, which can lead us to 
successful examples of inflation. A couple of them are shown in Fig.~2, 
where we plot $\Phi_{\rm i}^+$ as a function of the number of e-folds $N$.
It is clear then that a combination of the imaginary parts of the dilaton
$S$ and modulus $T$ fields, what we have denoted as $\Phi^+_{\rm i}$, can 
be a suitable inflaton in the context of heterotic M-theory. These are 
examples of the so-called natural inflation~\cite{Freese90}.

Let us finish this section by discussing very briefly the status of the 
moduli problem within these models. This arises when we have very 
weakly interacting particles with VEVs of the order of $M_{\rm P}$ and light
masses, of the order of $m_{3/2}$. If these relics decay, they must do it 
before nucleosynthesis in order not to ruin its predictions. This imposes a 
lower bound on their masses of $\sim 10$ TeV. On the other hand, if these
particles are stable, their oscillations should not overclose the Universe.
This sets an upper bound of $\sim 10^{-24}$ eV on their masses. In the weakly
coupled heterotic string case, where all moduli masses were of order $1$ TeV,
these bounds were obviously difficult to fulfil~\cite{Decar93}.

We have calculated these masses for the present models. In general, we obtain
that, along the $\Phi_{\rm r,i}^-$ directions, the corresponding masses are
of the order of $10^3 \; m_{3/2}$, well above the lower bound for decaying 
particles; along $\Phi_{\rm r}^+$ they are of the order of $10 \; m_{3/2}$, 
which may be just enough to save the bound, and for $\Phi_{\rm i}^+$ they are 
very small and dependent on how close $n_3$ is to $n_1$, $n_2$. For example, 
for ${\rm SU}(3)_{M=0} \times {\rm SU}(4)_{M=8} \times {\rm SU}(2)_{M=0}$ with
$n_1=n_2=1$ we find $m_{\Phi_{\rm i}^+} \sim  1.2 \times 10^{-3} \; m_{3/2}$ 
for $n_3=0.98$ and $m_{\Phi_{\rm i}^+} \sim  1.5 \times 10^{-4} \; m_{3/2}$ 
for $n_3=0.995$. There is therefore a conflict with the upper bound for 
stable particles, and the particle excess should be washed away with a 
period of thermal inflation~\cite{lyth95}.

\section{Conclusions}

We have studied the dynamics of the two typical M-theory moduli,
namely $\pi \rho$ and $V$ in terms of $S$, $T$. In the presence of 
non-perturbative corrections to the K\"ahler potential, and gaugino
condensation as the source of SUSY breaking, the scalar potential 
presents very interesting features.

$\bullet$ It depends on $S$, $T$ essentially through $f_a = S - n_a T/2$, 
motivating a redefinition of fields to $\Phi_{\rm r,i}^-$ and 
$\Phi_{\rm r,i}^+$.

$\bullet$ $\Phi_{\rm r,i}^-$ behave similarly to the dilaton of weakly 
coupled heterotic string theory: unsuitable as inflatons.

$\bullet$ $\Phi_{\rm r}^+$ would be flat in the absence of $K$: 
it has an inverse power-law potential, which gives very little intermediate 
inflation. 

$\bullet$ $\Phi_{\rm i}^+$ is totally flat for two condensates and can have 
a sinusoidal potential for three or more: very promising inflaton (many 
e-folds of natural inflation).

$\bullet$ There might be a  moduli problem associated to this
almost flat direction. These lighter moduli can be diluted with a small 
period of weak scale inflation.

\section*{Acknowledgments}
We thank the organizers for their hospitality. The work of TB is supported 
by PPARC.

\end{document}